\documentclass[prl,twocolumn,aps]{revtex4-1}
\usepackage{graphicx}% Include figure files
\usepackage{dcolumn}% Align table columns on decimal point
\usepackage{bm}% bold math
\usepackage{textcomp}
\usepackage{float}
\usepackage[colorlinks]{hyperref}
\usepackage{epstopdf}
\usepackage{amsmath}
\usepackage{amssymb}

\newcommand{\FP}{Fabry-P\'erot }
\begin{document}

\title{Single- and multi-mode Fabry-P\'erot interference in suspended graphene}

\author{Mika Oksanen$^1$}
\author{Andreas Uppstu$^2$}
\author{Antti Laitinen$^1$}
\author{Daniel J. Cox$^1$}
\author{Monica Craciun$^3$}
\author{Saverio Russo$^3$}
\author{Ari Harju$^2$}
\author{Pertti Hakonen$^1$}

\affiliation{$^1$Low Temperature Laboratory, Aalto University, P.O. Box 15100, FI-00076 AALTO, Finland \\
$^2$COMP Centre of Exellence and Helsinki Institute of Physics, Aalto University, P.O. Box 15100, FI-00076 AALTO, Finland \\
$^3$Centre for Graphene Science, School of Physics, University of Exeter, EX4 4QL Exeter, United Kingdom}

\date{\today}% It is always \today, today, but any date may be explicitly specified

\begin{abstract}

Phase coherence of charge carriers leads to electron-wave interference in ballistic mesoscopic conductors \cite{Dattabook}. In graphene, such Fabry-P\'erot-like interference has been observed \cite{Miao07,Du08,Young09,Rickhaus13}, but a detailed analysis has been complicated by the two-dimensional nature of conduction, which allows for complex interference patterns \cite{Gunlycke08,Shytov08,Muller09}. In this work, we have achieved high-quality Fabry-P\'erot interference in a suspended graphene device, both in conductance and in shot noise, and analyzed their structure using Fourier transform techniques. The Fourier analysis reveals two sets of overlapping, coexisting interferences, with the ratios of the diamonds being equal to the width to length ratio of the device. We attribute these sets to a unique coexistence of longitudinal and transverse resonances, with the longitudinal resonances originating from the bunching of modes with low transverse momentum. Furthermore, our results give insight into the renormalization of the Fermi velocity in suspended graphene samples, caused by unscreened many-body interactions \cite{Kotov12}.

\end{abstract}

\maketitle

The theoretical conditions for Fabry-P\'erot resonances have been analyzed in several recent works. Gunlycke and White \cite{Gunlycke08} showed that with non-perfect contacts, evenly spaced Fabry-P\'erot-like resonances should be observable in conductance measurements, with the periodicity being determined by the length of the sample and the velocity of the charge carriers. Such longitudinal resonances originate from simultaneous participation of modes in nonequivalent channels, facilitated by transversely quantized states with low transverse momentum and small energy separation. M\"{u}ller \emph{et al.} \cite{Muller09} emphasized that Fabry-P\'erot resonances should be observable at suitable gate voltage in experimental setups with metallic contacts, due to the presence of $pn$-junctions. Additionally, they theorized that in low temperature and with high-quality edges, resonances that correspond to single transverse modes should also be observable. Previously, equivalent single-mode resonances have been observed in single-wall carbon nanotubes (SWCNTs) \cite{Liang01}.

Experiments on suspended, exfoliated graphene have demonstrated mobilities exceeding 200 000 cm$^2$/Vs \cite{Bolotin08,Du08}, enabling experimental studies of ballistic transport in micron-sized graphene samples. Indeed, oscillations in the conductivity of graphene have been observed, indicating coherent transport and Fabry-P\'erot-like resonances \cite{Miao07,Heersche07,Du08,Rickhaus13}.  The interference patterns in graphene are similar to those in SWCNTs but, principally, more involved due to the two-dimensional nature of graphene, which leads to increased complexity due to the presence of a large number of conduction channels. Non-uniform spatial conditions, like charge puddles \cite{Zhang09} and flexural deformation \cite{Castro10}, complicate the situation even further. As a consequence, no coherent picture of Fabry-P\'erot interference in suspended graphene devices has emerged by now.

In this work, we employ conductance and shot noise measurements to analyze transport resonances with characteristic features of Fabry-P\'erot  interference. Our results show resonances that are attributable to both transverse (single-mode) and longitudinal (multi-mode) interference. Moreover, the suspended region of the sample yields a Fermi velocity $v_F$ of the order of $ 2.4-2.8 \times 10^6$~m/s at a charge density of $ n \sim 1-2 \times 10^{10}$~cm$^{-2}$. Although the exact value of $v_F$ may be overestimated because of anomalies like non-uniform charge distribution \cite{Grushina13}, both theoretical predictions and experimental observations have suggested that in suspended graphene, many-body interactions may renormalize $v_F$ to a significantly higher value than found in graphene on a substrate \cite{Elias11, Kotov12}.

The locations of Fabry-P\'erot resonances in a graphene device of length $L$ and width $W$ are obtained by modeling the device as a quantum dot, which gives  \cite{Muller09, Gunlycke08}
\begin{equation}
E_{q_L,q_W} \cong  \pm \sqrt {E_L^2 {{\left( {q_L +\delta_L} \right)}^2} + {E_W^2}{{\left( {q_W + \delta_W } \right)}^2}},
\label{energymn}
\end{equation}
where $E_L \equiv hv_F/2L$ and $E_W \equiv hv_F/2W$. The values of the constants $\delta_L$ and $\delta_W$ depend on the details of the interfaces or edges, such as the occurrence of a Berry phase \cite{Gunlycke08}, and the quantum numbers $q_L$ and $q_W$ label longitudinal and transverse modes, respectively. According to Eq. (\ref{energymn}), in a device with $W/L \gg 1$, modes with low $q_W$ are bunched together at integer values of $E_L$, thus signifying a multichannel Fabry-P\'erot interference phenomenon. On the other hand, at fixed $q_L$, modes with large $q_W$ are spaced roughly as $E_W$. The widths of the resonances are approximately given by $\delta \mu \cong \sqrt{2} (q_L \pi)^2 E_L^3/\mu^2$ \cite{Muller09}, with the sharpest peaks corresponding to low $q_L$ and high $\mu$.

\begin{figure*}[h]
\centering
\includegraphics[width=\textwidth]{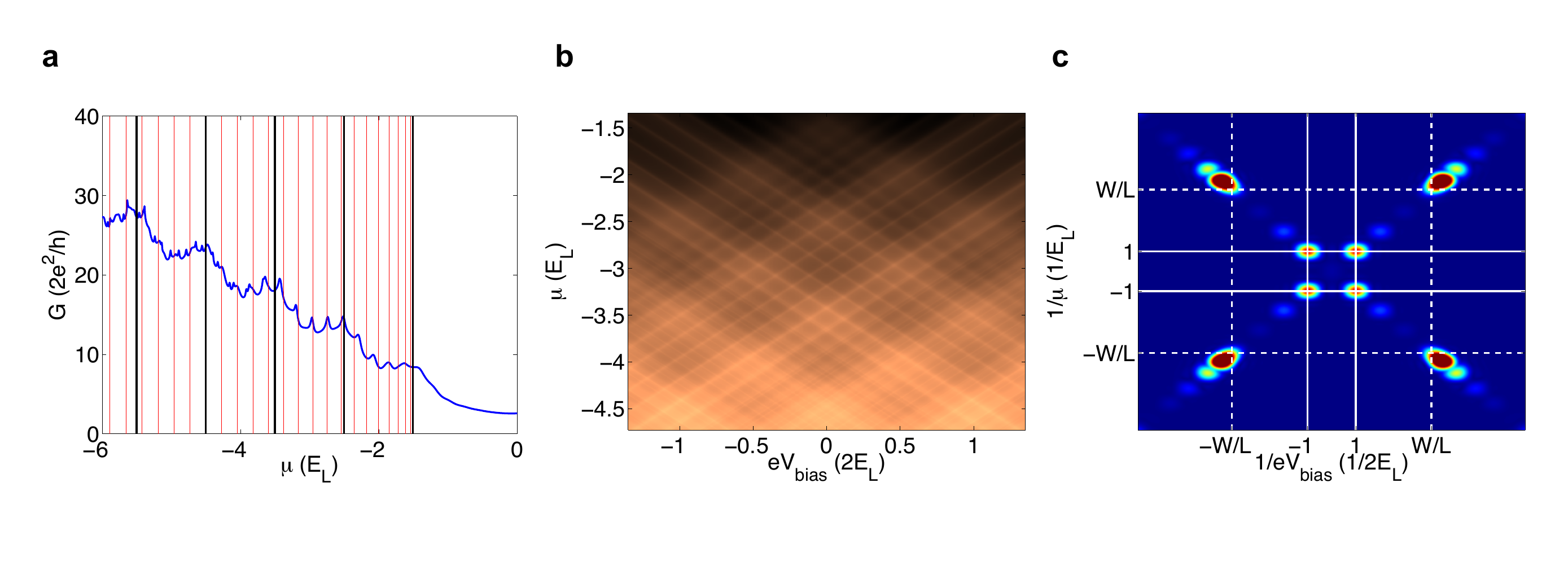}
\caption{(a) Simulated conductance curve showing the presence of fast and slow oscillations, corresponding to transverse and longitudinal Fabry-P\'erot interference, respectively. The bars indicate resonances given by Eq. (\ref{energymn}), with the red bars corresponding to $q_L=1$ and arbitrary $q_W$, and the black bars to $q_W=0$ and arbitrary $q_L$. (b) Simulated map of the differential conductance $G_d$. (c) Fourier transform of $\partial G_d / \partial \mu$, highlighting the periodicities of the two sets of resonances.}
\label{f1}
\end{figure*}

We have illustrated the appearance of the resonances through a tight-binding transport simulation of a graphene device with a $W/L$-ratio of $\sim 4$. Metallic contacts are modeled by semi-infinite graphene leads at a constant doping of 1~eV, and enhanced scattering at the contact-device interfaces is simulated by reducing the hopping values across the interfaces by 30\%. Fig. \ref{f1}a shows a zoom-in on the $pn$-side of the resulting conductance curve, which clearly contains both fast and slow oscillations. The locations of some of the resonances given by Eq.~(\ref{energymn}) have been indicated with bars. The red bars correspond to fast transverse resonances with varying $q_W$ but $q_L$ set to one, while approximate positions for the slow longitudinal resonances are given by the black bars, which correspond to $q_W=0$ and varying $q_L$. Fig. \ref{f1}b shows a  conductance map that highlights the presence of two sets of diamonds. The resonances can be analyzed quantitatively by first differentiating the map with respect to $\mu$, and then performing a Fourier transform. The slow periodicity of $E_L$ is clearly visible in the Fourier transformed map shown in Fig. \ref{f1}c, while the periodicity of the fast resonances is slightly lower than $E_W$, due to resonances that correspond to relatively low $q_W$.

\begin{figure*}[h]
\centering
\includegraphics[width=\textwidth]{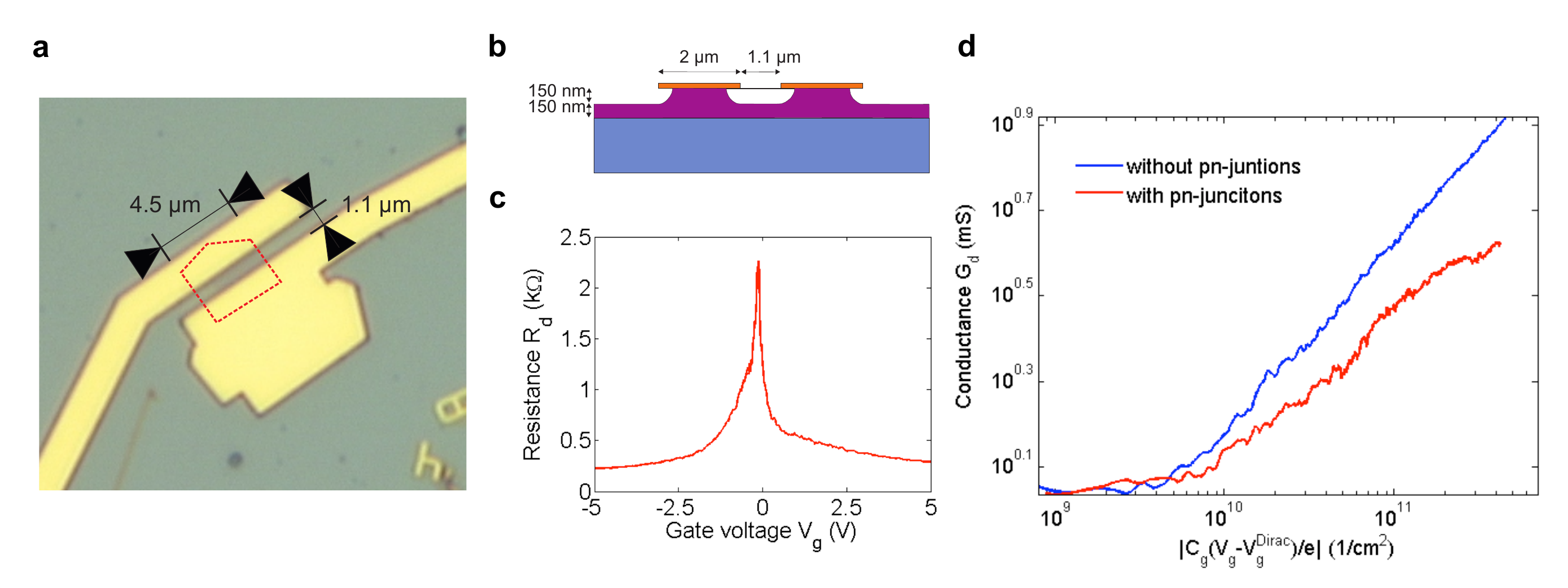}
\caption{(a) Optical image of the sample, where the dashed line indicates the position of the graphene flake. (b) Schematic structure of the suspended graphene device. The HF etching depth of SiO$_2$ is limited to about 150 nm. (c) Resistance of the sample as a function of the gate voltage up to $\pm 5$ V, which corresponds to a charge density range up to $\pm 15 \times 10^{10}$ cm$^{-2}$. (d) Conductance vs. $|C_g (V_g-V_g^\text{Dirac})/e|$ on a log-log scale. The solid lines yield an estimate of $n_r \simeq 6 \times 10^{10}$ cm$^{-2}$ for the residual charge density.}
\label{f2}
\end{figure*}

Our experiments were performed at a temperature of 50 mK using a suspended graphene sample with length $L=1.1$ $\mu$m and width $W=4.5$ $\mu$m. The experimental setup is shown in Figures \ref{f2}a and b.  As the metal contacts of the setup cause $n$-doping of the underlying graphene, at negative gate voltage $pn$-junctions are formed close to the contacts. This causes increased reflectivity \cite{Cheianov06, Cheianov07}, which assists in the formation of \FP resonances. The measured minimum conductivity falls below $\frac{4e^2}{h}$ approaching theoretical \cite{Tworzydlo06} minimum conductivity $\sigma _\text{min} = \frac{4e^2}{\pi h}$ for high aspect ratio samples, the actual maximum resistance of the sample was 2.2 k$\Omega$, as shown in Fig. \ref{f2}c. The charge density on the device at zero bias is $|n_0|=\sqrt{C_g^2 (V_g-V_g^\text{Dirac})^2/e^2+n_r^2}$, where $C_g$ is the capacitance per unit area of the gate setup, $V_g^\text{Dirac}$ the location of the Dirac point, and $n_r$ the residual charge density due to impurities. $n_r$ can be determined using the $\log(G)$ vs $\log(|C_g(V_g-V_g^\text{Dirac})/e|)$ plot shown in Fig. \ref{f2}d, based on the point where the linear behavior levels to a constant value at low $\log(|C_g(V_g-V_g^\text{Dirac})/e|)$. The estimation leads to a value of $n_r \sim 6 \times 10^{9}$ cm$^{-2}$ 

\begin{figure*}[h]
\centering
\includegraphics[width=\textwidth]{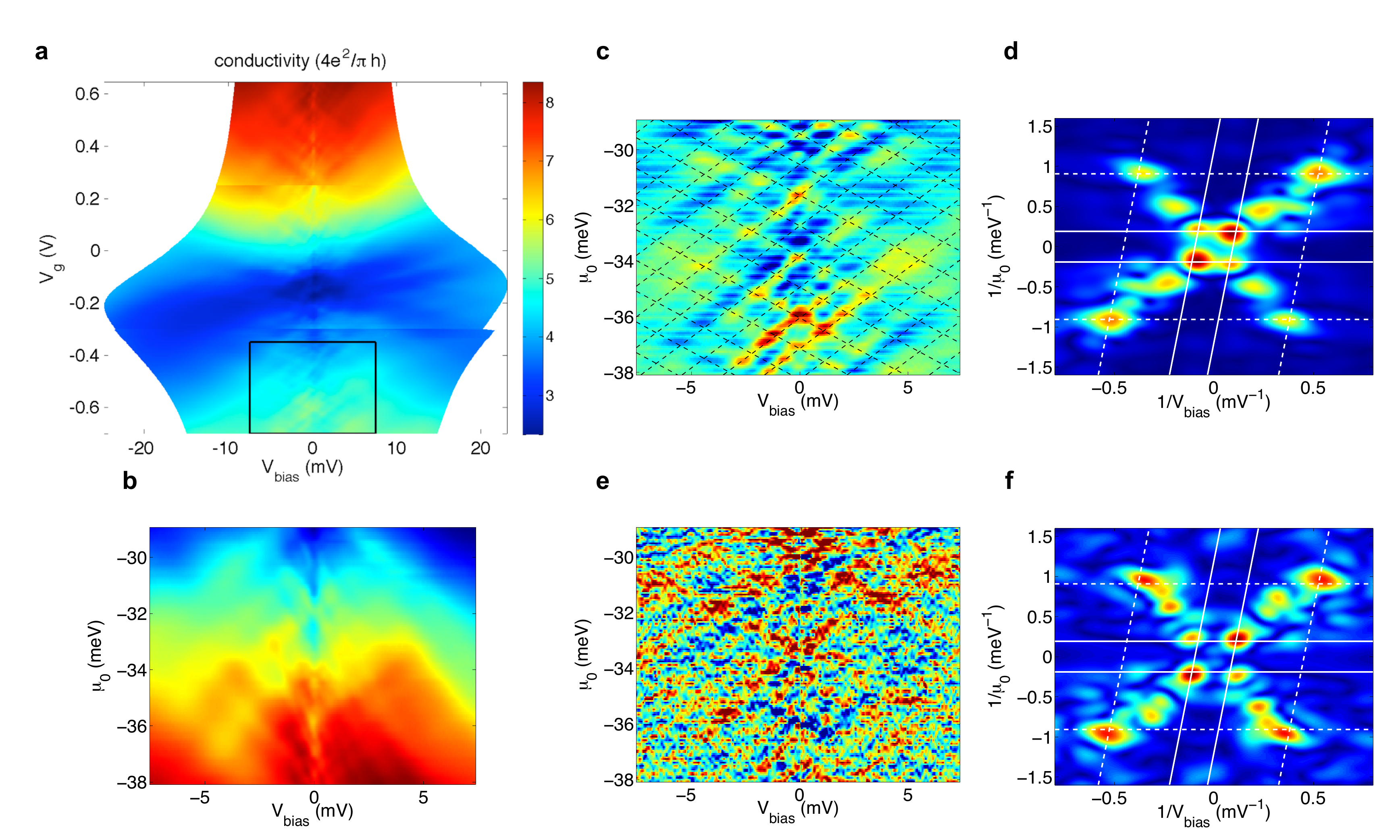}
\caption{(a) Measured differential conductance map of the graphene device. (b) A zoom-in on the boxed region in (a), with the gate voltage converted into chemical potential. (c) The conductance map in (b) differentiated with respect to the chemical potential. The dashed lines are a fit of resonances with periodicity $E_W$, using $v_F=2.4 \times 10^6$ m/s. (d) Fourier transform of (c). The solid lines are fit to the longitudinal interference and the dashed to the transverse interference, using the same $v_F$ as in (c). (e) Differential Fano-factor differentiated with respect to the chemical potential. (f) Fourier transform of (e), with the solid and dashed lines being the same as those shown in (d).}
\label{f3}
\end{figure*}

In the scans over the bias voltage $V_b$ and gate voltage $V_g$, a Fabry-P\'erot pattern is observed both in conductance and in shot noise. To quantitatively study the experimental map of the differential conductance $G_d \equiv dI/dV$, shown in Fig. \ref{f3}a, we convert the gate voltage into a low-bias chemical potential $\mu_0$. Using the density of states of graphene, the charge density can be converted into a zero-bias chemical potential through $\mu_0=\text{sgn}(n_0) \hbar v_F \sqrt{\pi |n_0|} $. Fig. \ref{f3}b shows a  zoom-in of the converted data at a charge density of $ n_0=1.1-1.8 \times 10^{10}$ cm$^{-2}$, indicated by the box in Fig. \ref{f3}a. A diamond pattern appears if the conductance is differentiated with respect to the zero-bias chemical potential. The differentiated data, displayed in Fig. \ref{f3}c, indicates that there is a periodic modulation in $G_d$, especially visible well away from the Dirac point, which is located around $V_g=-0.15$ V. A good fit with the experimental data can be obtained, if $C_g$ is set to the value given by a plane capacitor model of the gate setup, i.e. 47 aF/$\mu$m$^2$, and $n_r$ to $9 \times 10^9 \text{cm}^{-2}$, which is close to the value estimated using Fig. \ref{f2}d.

The diamonds are not completely symmetric, which may be explained through slightly asymmetrical contacts. Besides causing a shift of the chemical potential of one of the contacts (the other one is grounded), a finite bias voltage also causes a shift of the chemical potential of the graphene device. Here, we assume that this shift is linear, i.e. $\mu=\mu_{0}+xeV_b$, where $\mu$ is the chemical potential of the device at arbitrary bias. This results in two sets of evenly spaced resonances as a function of bias voltage at a fixed gate voltage. One set is spaced as $\Delta E/x$ and the other as $\Delta E / (1-x)$. In the $G_d$ map, a set of diamonds with aspect ratio two is then formed, being completely symmetrical when $x=0.5$. With the current experimental data, however, a best fit is achieved with $x=0.58$. Additionally, by setting $W=4.5~\mu$m and $v_F=2.4 \times 10^6$ m/s, we obtain the dashed lines shown in Fig. \ref{f3}c. The high Fermi velocity, well above values measured on a SiO$_2$ substrate \cite{Novoselov05, Zhang05}, is consistent with measurements of the cyclotron mass in freestanding graphene \cite{Elias11}, which have indicated that $v_F$  may be between 2 and $3 \times 10^6$ m/s at similar charge density. This effect is thought to be caused by unscreened electron-electron interactions occuring in suspended graphene \cite{Kotov12}. As in the simulation, Fourier analysis may be applied to reveal the presence of multi-mode resonances spaced as $E_L$. In the Fourier-transformed plot, shown in Fig. \ref{f3}d, a strong resonance is indeed found at a periodicity that corresponds to a value close to $L=1.1~\mu$m. The transverse resonances are also clearly visible in the Fourier-transformed plot, as indicated by the dashed lines, which correspond to the periodicity of the dashed diamonds shown in Fig. \ref{f3}c. If we assume that the average periodicity of the fast resonances is slightly less than $E_W$, as in the simulation, we obtain a slightly higher $v_F$ of roughly $2.8 \times 10^6$ m/s.

Shot noise yields additional information on the distribution of transmission channels as well as interaction effects \cite{Danneau08a,Wu07}. Shot noise can be quantified through the differential Fano-factor, which is defined as $\mathcal{F}_d\equiv (1/2e) dS/dI$, where $S$ is the correlation function of the current fluctuations $\delta I(t)$, i.e. $S=\int dt \langle \delta I(t) \delta I(0)  \rangle$. Our measured results on shot noise are illustrated in Fig. \ref{f3}e, which depicts the derivative $\partial \mathcal{F}_d /\partial \mu$ as a function of $\mu_0$ and $V_b$ over the same area as in Fig. \ref{f3}b. Although the resonances in Fig. \ref{f3}e are not straightforward to interpret, the Fourier transform shown in Fig. \ref{f3}f reveals two clear sets of resonances with the same periodicities as the conductance data, with the solid and dashed lines corresponding to exactly the same values as in Fig. \ref{f3}d. The Fourier transforms of both the $\partial G_d/ \partial  \mu$ and $\partial \mathcal{F}_d/\partial \mu$ maps contain additional spots between the innermost and outermost resonances. These might arise from the presence of additional scattering surfaces in the experimental setup, or they might simply be harmonic multiplets of the lowest resonance, as indicated by the simulation.

To summarize, we have measured the conductance of a suspended graphene sample with a high aspect ratio. Through Fourier analysis techniques, the differential conductance and the differential shot noise reveal two sets of clear resonances, which may be attributed to single-mode transverse and multi-mode longitudinal resonance. Although theoretically anticipated, their coexistence has previously not been observed. The analysis of the interference pattern  yields a Fermi velocity that is significantly higher than the one usually reported for graphene on a substrate, thus indicating that close to the Dirac point, unscreened many-body interactions may be significant in suspended graphene.

\section{METHODS}

The experiments were performed on a Bluefors LD250 dilution refrigerator. Before the measurements, the two-lead sample was annealed by passing a current of 1.1 mA through it (using a voltage of nearly 0.9 V over the sample). This resulted in an almost neutral, high-quality sample with the charge neutrality point located at $V_g^\text{Dirac} = -0.15$ V. The field-effect mobility was estimated to be $\mu >10^5$ cm$^2$/Vs at charge densities of $n < 2.5 \times 10^{10}$ cm$^{-2}$. An estimate for the contact resistance, $R_c \lesssim 100$ $\Omega$, was obtained based on the measured Fano-factor $\mathcal{F} = 0.2$ at $n \sim 1-2 \cdot 10^{10}$ cm$^{-2}$, by assuming that the total shot noise is induced by a tunneling contact with $\mathcal{F}= 1$.

The main features of the measurement setup are described in Ref. \onlinecite{Danneau08b}. We measured both the differential conductance and the differential Fano-factor. Lines connected to the graphene device (source and drain) were connected to bias-Ts separating low frequency signals (less than 1 kHz) and the high-frequency noise measurement signals above 600 MHz, DC biasing via a 0.5 M$\Omega$ resistor was employed. The high frequency line on the source side was terminated by a 50 $\Omega$ shunt and the line on the drain side (shot noise measurement line) was connected to a low noise amplifier through a circulator, preventing the amplifier noise interfering with the sample. A tunnel junction and a microwave switch were employed for calibration of the noise spectrometer. Standard lock-in techniques were employed for measurements of the low frequency conductance (around 35 Hz), the same excitation could also be employed to measure differential shot noise by directing the rectified output of the noise spectrometer to a lock-in amplifier.

Fourier transforms were performed after converting the gate voltage to a zero-bias chemical potential. A Hanning window and padding with zeroes were employed in order to avoid artifacts in the the Fourier transform, computed using FFT techniques. We determined the Fermi velocity by fitting $E_L$ and $E_W$ to the bias values of the resonances in the Fourier transform. An estimate for $n_r$ could be obtained by requiring a fit to the theoretical values also in the direction of the chemical potential and setting $C_g$ to the value estimated by a plane capacitor model.

\bibliographystyle{naturemag}
%\bibliography{FPrefs.bib}

\begin{thebibliography}{10}
\expandafter\ifx\csname url\endcsname\relax
  \def\url#1{\texttt{#1}}\fi
\expandafter\ifx\csname urlprefix\endcsname\relax\def\urlprefix{URL }\fi
\providecommand{\bibinfo}[2]{#2}
\providecommand{\eprint}[2][]{\url{#2}}

\bibitem{Dattabook}
\bibinfo{author}{Datta, S.}
\newblock \emph{\bibinfo{title}{{Electronic Transport in Mesoscopic Systems}}}
  (\bibinfo{publisher}{Cambridge University Press}, \bibinfo{year}{1997}).

\bibitem{Miao07}
\bibinfo{author}{Miao, F.} \emph{et~al.}
\newblock \bibinfo{title}{{Phase-Coherent Transport in Graphene Quantum
  Billiards}}.
\newblock \emph{\bibinfo{journal}{Science}} \textbf{\bibinfo{volume}{317}},
  \bibinfo{pages}{1530--1533} (\bibinfo{year}{2007}).

\bibitem{Du08}
\bibinfo{author}{Du, X.}, \bibinfo{author}{Skachko, I.},
  \bibinfo{author}{Barker, A.} \& \bibinfo{author}{Andrei, E.~Y.}
\newblock \bibinfo{title}{{Approaching ballistic transport in suspended
  graphene}}.
\newblock \emph{\bibinfo{journal}{Nature Nanotechnology}}
  \textbf{\bibinfo{volume}{3}}, \bibinfo{pages}{491--495}
  (\bibinfo{year}{2008}).

\bibitem{Young09}
\bibinfo{author}{Young, A.~F.} \& \bibinfo{author}{Kim, P.}
\newblock \bibinfo{title}{{Quantum interference and Klein tunnelling in
  graphene heterojunctions}}.
\newblock \emph{\bibinfo{journal}{Nature Physics}}
  \textbf{\bibinfo{volume}{5}}, \bibinfo{pages}{222--226}
  (\bibinfo{year}{2009}).

\bibitem{Rickhaus13}
\bibinfo{author}{{Rickhaus}, P.} \emph{et~al.}
\newblock \bibinfo{title}{{Ballistic interferences in suspended graphene}}.
\newblock \emph{\bibinfo{journal}{ArXiv e-prints}}  (\bibinfo{year}{2013}).
\newblock \eprint{1304.6590}.

\bibitem{Gunlycke08}
\bibinfo{author}{Gunlycke, D.} \& \bibinfo{author}{White, C.~T.}
\newblock \bibinfo{title}{{Graphene interferometer}}.
\newblock \emph{\bibinfo{journal}{Applied Physics Letters}}
  \textbf{\bibinfo{volume}{93}}, \bibinfo{pages}{2106} (\bibinfo{year}{2008}).

\bibitem{Shytov08}
\bibinfo{author}{Shytov, A.~V.}, \bibinfo{author}{Rudner, M.~S.} \&
  \bibinfo{author}{Levitov, L.~S.}
\newblock \bibinfo{title}{{Klein Backscattering and Fabry-P{\'e}rot
  Interference in Graphene Heterojunctions}}.
\newblock \emph{\bibinfo{journal}{Physical Review Letters}}
  \textbf{\bibinfo{volume}{101}}, \bibinfo{pages}{156804}
  (\bibinfo{year}{2008}).

\bibitem{Muller09}
\bibinfo{author}{M{\"u}ller, M.}, \bibinfo{author}{Br{\"a}uninger, M.} \&
  \bibinfo{author}{Trauzettel, B.}
\newblock \bibinfo{title}{{Temperature Dependence of the Conductivity of
  Ballistic Graphene}}.
\newblock \emph{\bibinfo{journal}{Physical Review Letters}}
  \textbf{\bibinfo{volume}{103}}, \bibinfo{pages}{196801}
  (\bibinfo{year}{2009}).

\bibitem{Kotov12}
\bibinfo{author}{Kotov, V.~N.}, \bibinfo{author}{Uchoa, B.},
  \bibinfo{author}{Pereira, V.~M.}, \bibinfo{author}{Guinea, F.} \&
  \bibinfo{author}{Castro~Neto, A.~H.}
\newblock \bibinfo{title}{{Electron-Electron Interactions in Graphene: Current
  Status and Perspectives}}.
\newblock \emph{\bibinfo{journal}{Reviews of Modern Physics}}
  \textbf{\bibinfo{volume}{84}}, \bibinfo{pages}{1067--1125}
  (\bibinfo{year}{2012}).

\bibitem{Liang01}
\bibinfo{author}{Liang, W.} \emph{et~al.}
\newblock \bibinfo{title}{{Fabry-Perot interference in a nanotube electron
  waveguide}}.
\newblock \emph{\bibinfo{journal}{Nature}} \textbf{\bibinfo{volume}{411}},
  \bibinfo{pages}{665--669} (\bibinfo{year}{2001}).

\bibitem{Bolotin08}
\bibinfo{author}{Bolotin, K.~I.}, \bibinfo{author}{Sikes, K.~J.},
  \bibinfo{author}{Hone, J.}, \bibinfo{author}{Stormer, H.~L.} \&
  \bibinfo{author}{Kim, P.}
\newblock \bibinfo{title}{Temperature-dependent transport in suspended
  graphene}.
\newblock \emph{\bibinfo{journal}{Physical Review Letters}}
  \textbf{\bibinfo{volume}{101}}, \bibinfo{pages}{096802}
  (\bibinfo{year}{2008}).

\bibitem{Heersche07}
\bibinfo{author}{Heersche, H.~B.}, \bibinfo{author}{Jarillo-Herrero, P.},
  \bibinfo{author}{Oostinga, J.~B.}, \bibinfo{author}{Vandersypen, L. M.~K.} \&
  \bibinfo{author}{Morpurgo, A.~F.}
\newblock \bibinfo{title}{{Bipolar supercurrent in graphene}}.
\newblock \emph{\bibinfo{journal}{Nature}} \textbf{\bibinfo{volume}{446}},
  \bibinfo{pages}{56--59} (\bibinfo{year}{2007}).

\bibitem{Zhang09}
\bibinfo{author}{Zhang, Y.}, \bibinfo{author}{Brar, V.~W.},
  \bibinfo{author}{Girit, C.}, \bibinfo{author}{Zettl, A.} \&
  \bibinfo{author}{Crommie, M.~F.}
\newblock \bibinfo{title}{Origin of spatial charge inhomogeneity in graphene}.
\newblock \emph{\bibinfo{journal}{Nature Physics}}
  \textbf{\bibinfo{volume}{5}}, \bibinfo{pages}{722--726}
  (\bibinfo{year}{2009}).

\bibitem{Castro10}
\bibinfo{author}{Castro, E.~V.} \emph{et~al.}
\newblock \bibinfo{title}{{Limits on Charge Carrier Mobility in Suspended
  Graphene due to Flexural Phonons}}.
\newblock \emph{\bibinfo{journal}{Physical Review Letters}}
  \textbf{\bibinfo{volume}{105}}, \bibinfo{pages}{266601}
  (\bibinfo{year}{2010}).

\bibitem{Grushina13}
\bibinfo{author}{Grushina, A.~L.}, \bibinfo{author}{Ki, D.-K.} \&
  \bibinfo{author}{Morpurgo, A.~F.}
\newblock \bibinfo{title}{A ballistic pn junction in suspended graphene with
  split bottom gates}.
\newblock \emph{\bibinfo{journal}{Applied Physics Letters}}
  \textbf{\bibinfo{volume}{102}}, \bibinfo{pages}{223102}
  (\bibinfo{year}{2013}).

\bibitem{Elias11}
\bibinfo{author}{Elias, D.~C.} \emph{et~al.}
\newblock \bibinfo{title}{{Dirac cones reshaped by interaction effects in
  suspended graphene}}.
\newblock \emph{\bibinfo{journal}{Nature Physics}}
  \textbf{\bibinfo{volume}{7}}, \bibinfo{pages}{701--704}
  (\bibinfo{year}{2011}).

\bibitem{Cheianov06}
\bibinfo{author}{Cheianov, V.~V.} \& \bibinfo{author}{Fal'ko, V.~I.}
\newblock \bibinfo{title}{{Selective transmission of Dirac electrons and
  ballistic magnetoresistance of n-p junctions in graphene}}.
\newblock \emph{\bibinfo{journal}{Physical Review B}}
  \textbf{\bibinfo{volume}{74}}, \bibinfo{pages}{41403} (\bibinfo{year}{2006}).

\bibitem{Cheianov07}
\bibinfo{author}{Cheianov, V.~V.}, \bibinfo{author}{Fal~ko, V.} \&
  \bibinfo{author}{Altshuler, B.~L.}
\newblock \bibinfo{title}{{The Focusing of Electron Flow and a Veselago Lens in
  Graphene p-n Junctions}}.
\newblock \emph{\bibinfo{journal}{Science}} \textbf{\bibinfo{volume}{315}},
  \bibinfo{pages}{1252} (\bibinfo{year}{2007}).

\bibitem{Tworzydlo06}
\bibinfo{author}{Tworzyd{\l}o, J.}, \bibinfo{author}{Trauzettel, B.},
  \bibinfo{author}{Titov, M.}, \bibinfo{author}{Rycerz, A.} \&
  \bibinfo{author}{Beenakker, C.}
\newblock \bibinfo{title}{{Sub-Poissonian Shot Noise in Graphene}}.
\newblock \emph{\bibinfo{journal}{Physical Review Letters}}
  \textbf{\bibinfo{volume}{96}}, \bibinfo{pages}{246802}
  (\bibinfo{year}{2006}).

\bibitem{Novoselov05}
\bibinfo{author}{Novoselov, K.~S.} \emph{et~al.}
\newblock \bibinfo{title}{{Two-dimensional gas of massless Dirac fermions in
  graphene}}.
\newblock \emph{\bibinfo{journal}{Nature}} \textbf{\bibinfo{volume}{438}},
  \bibinfo{pages}{197--200} (\bibinfo{year}{2005}).

\bibitem{Zhang05}
\bibinfo{author}{Zhang, Y.}, \bibinfo{author}{Tan, Y.-W.},
  \bibinfo{author}{Stormer, H.~L.} \& \bibinfo{author}{Kim, P.}
\newblock \bibinfo{title}{{Experimental observation of the quantum Hall effect
  and Berry's phase in graphene}}.
\newblock \emph{\bibinfo{journal}{Nature}} \textbf{\bibinfo{volume}{438}},
  \bibinfo{pages}{201--204} (\bibinfo{year}{2005}).

\bibitem{Danneau08a}
\bibinfo{author}{Danneau, R.} \emph{et~al.}
\newblock \bibinfo{title}{{Shot Noise in Ballistic Graphene}}.
\newblock \emph{\bibinfo{journal}{Physical Review Letters}}
  \textbf{\bibinfo{volume}{100}}, \bibinfo{pages}{196802}
  (\bibinfo{year}{2008}).

\bibitem{Wu07}
\bibinfo{author}{Wu, F.} \emph{et~al.}
\newblock \bibinfo{title}{{Shot Noise with Interaction Effects in Single-Walled
  Carbon Nanotubes}}.
\newblock \emph{\bibinfo{journal}{Physical Review Letters}}
  \textbf{\bibinfo{volume}{99}}, \bibinfo{pages}{156803}
  (\bibinfo{year}{2007}).

\bibitem{Danneau08b}
\bibinfo{author}{Danneau, R.} \emph{et~al.}
\newblock \bibinfo{title}{{Evanescent Wave Transport and Shot Noise in
  Graphene: Ballistic Regime and Effect of Disorder}}.
\newblock \emph{\bibinfo{journal}{Journal of Low Temperature Physics}}
  \textbf{\bibinfo{volume}{153}}, \bibinfo{pages}{374--392}
  (\bibinfo{year}{2008}).

\end{thebibliography}

\end{document}